\documentclass[conference]{IEEEtran}
\IEEEoverridecommandlockouts
\usepackage{cite}
\usepackage{amsmath,amssymb,amsfonts}
\usepackage{algorithmic}
\usepackage{algorithm2e,float}
\usepackage{graphicx}
\usepackage{subfigure}
\usepackage{multirow}
\usepackage{textcomp}
\usepackage{xcolor}
\usepackage{hyperref}
\usepackage{array}
\usepackage{adjustbox}
\usepackage{pgfplots}
\pgfplotsset{compat=1.17}
\usepackage{float}
\def\BibTeX{{\rm B\kern-.05em{\sc i\kern-.025em b}\kern-.08em
    T\kern-.1667em\lower.7ex\hbox{E}\kern-.125emX}}
\begin{document}

\title{LogShield: A Transformer-based APT Detection System Leveraging Self-Attention 
}

\author{
  \IEEEauthorblockN{\large{\textbf{Sihat Afnan\IEEEauthorrefmark{2}\textsuperscript{*}, Mushtari Sadia\IEEEauthorrefmark{2}\textsuperscript{*}\thanks{* Both authors contributed equally.}, Shahrear Iqbal\IEEEauthorrefmark{3}, and Anindya Iqbal\IEEEauthorrefmark{2}}}}
  \IEEEauthorblockA{
    \IEEEauthorrefmark{2}Bangladesh University of Engineering and Technology\\
    \IEEEauthorrefmark{3}National Research Council Canada
  }
}

\maketitle

\begin{abstract}
Cyber attack detection is a critical task in ensuring the security and reliability of a system. Attacks are often identified using system and network logs, a primary data source across computer systems. There have been significant prior works that utilize provenance graphs and machine learning techniques to detect attacks, specifically Advanced Persistent Threats (APT) which are very difficult to detect due to their low and slow approach. Lately, there has been emerging research where transformer-based language models are being used to detect various types of attacks from system logs. However, no such attempts have been made in the case of APTs. In addition, existing state-of-the-art techniques that use system provenance graphs, lack a data processing framework generalized across datasets for optimal performance. For exploring the effectiveness of transformer-based language models as well as mitigating the aforementioned limitation, this paper proposes LogShield, a framework designed to detect APT attack patterns leveraging the power of self-attention in transformers. We incorporate customized embedding layers to effectively capture the context of event sequences derived from provenance graphs. While acknowledging the computational overhead associated with training transformer networks, our framework surpasses existing LSTM and Language models regarding APT detection performance. We integrated the model parameters and training procedure from the RoBERTa model and conducted extensive experiments on well-known APT datasets (DARPA OpTC and DARPA TC E3). Our framework achieved superior F1 scores of 98\% and 95\% on the two datasets respectively, surpassing the F1 scores of 96\% and 94\% obtained by LSTM models. Our findings suggest that LogShield's performance benefits from larger datasets and demonstrates its potential for generalization across diverse domains. These findings contribute to the advancement of APT attack detection methods and underscore the significance of transformer-based architectures in addressing security challenges in computer systems.

\end{abstract}


\section{Introduction}
The dynamic nature of cyber-attacks, especially advanced persistent threats or APTs, demands a proactive approach to safeguarding sensitive information and critical infrastructures. APTs are characterized by their stealth nature and ability to remain undetected for long periods of time. They are typically targeted attacks, meaning that they are directed at specific organizations or individuals for the purpose of stealing sensitive data or disrupting operations. These attacks are often carried out by nation-states or other well-funded and skilled cybercrime groups, and they may involve the use of custom malware and other advanced tactics. According to APT trends report Q3 2022 \cite{apttrendsreport2022}, APT campaigns are very widely spread geographically, expanding into Europe, the US, Korea, Brazil, the Middle East and various parts of Asia.

According to recent research \cite{b7,xiong2020conan}, data provenance may be a reliable data source for APT detection. Data provenance utilizes system logs by representing all system events as a directed acyclic graph (DAG) that describes information flow between system subjects (processes) and objects (files and sockets). It connects causally-related events in the graph, even when those events are separated by a long time period. Thus, even though systems under attack usually behave similarly to unattacked systems, the richer contextual information in provenance allows for better separation of benign and malicious events.

In recent years, as demonstrated by BERT or GPT-3, language models have been seen to receive state of the art results in various fields \cite{brown2020language}. One possible domain of their application can be APT detection. Numerous types of system data such as system logs or event traces can be structured as texts and viewed as a separate language. There has already been notable work in this field. In Yesir et al. \cite{yesir2021malware}, malware classification was performed using BERT and fastText and a satisfactory result was achieved. More recently, in Alkhatib et al. \cite{alkhatib2022can}, a framework called CAN-BERT was developed to perform intrusion detection using BERT . In Guo et al \cite{guo2021logbert}, a self-supervised framework for log anomaly detection based on Transformers was developed which outperformed state-of-the-art approaches for anomaly detection, by experimenting on three log datasets.

The motivation behind using language models to detect malicious event traces lies in the fact that language models are designed to take into account the sequence of words in a sentence. Therefore, sequential log data, which refers to a log or record of events that are recorded in a specific order, can be processed by language models. This is important because the sequence of events can convey the meaning and context of a process which would be lost if the event traces were rearranged. For example, consider a web browser (e.g., chrome.exe) that opens a PDF file by triggering a PDF viewer (e.g., acrobat.exe). This is a benign action. The opposite sequence, where the PDF viewer triggers the web browser, could potentially be considered a malicious action, as it might indicate an attempt by the PDF file to execute code within the web browser. Another example to consider is a chat program (e.g., slack.exe) opening a chat window by triggering a web browser (e.g., firefox.exe). In contrast, if the web browser triggers the chat process, it could be indicative of a malicious event.

Another reason behind using language models for APT  detection is that the self-attention mechanism of transformer-based language models allows each word, or in the case of logs, each event in a sequence to interact with all other events, fostering a holistic understanding of the entire sequence. This attention mechanism assigns different weights to different events based on their relevance, dynamically adjusting each event's importance as the model processes the sequence. In the context of log data, this translates to the model being able to focus on critical events that might indicate malicious activities while considering the broader context of the entire log sequence. 

As deep learning models, particularly recurrent neural networks (RNNs) can represent sequential data, they are now frequently used for log anomaly detection. Nevertheless, there are still certain limitations of RNNs to detect sequences in log data. Traditional RNNs suffer from vanishing and exploding gradient problems, which restrict their ability to capture long-range dependencies effectively. In contrast, the self-attention mechanism in Transformer models is not subject to these limitations. This allows Transformers to capture both local and global dependencies efficiently, making them well-suited for handling the complex and diverse patterns found in log data.

In this paper, we propose a generalized framework to process log data collected from systems under APT attack and detect the attack leveraging self-attention. We have adopted a system provenance-based approach to generate event traces. Our findings indicate that existing deep learning techniques employed for APT detection experience performance degradation compared to transformer-based language models, as the volume and length of event sequences grow. However, we also observed that transformer-based models face challenges when provided with raw log data, as they fail to capture certain crucial aspects indicating malicious events. To address this issue, our framework focuses on extracting specific, essential information from the log data enabling language models to effectively capture these critical indicators. By doing so, our approach ensures that transformer-based models perform optimally in detecting attacks, surpassing the limitations of existing methods.

% A host based provenance graph was constructed utilizing the parent child relationship of event logs. From random walks on the graph, event traces were generated as well. Starting from a random node, an event trace was generated until leaf node was found. The event traces not only comes in handy as modeling input to transformer based language models but bears significant insight about the events, the temporal information of the events or object-action pairs that differentiate malicious event logs from the benign data. It can lead us to understanding the context in which a vulnerability is exploited, helps to gain valuable insights into the tactics and methods used by attackers, as well as identifying important trends in attack techniques.

We have used the DARPA OpTC \cite{optcrepository} and the DARPA TC E3 \cite{darpa-tc} datasets to validate our approach. The DARPA TC E3 dataset is obtained from the third engagement exercise of the DARPA Transparent Computing (TC) program, which involved benign data generation initially, followed by simulation of APT attack behaviors during specific weekday hours while maintaining continuous benign background traffic. The DARPA OpTC dataset is the largest publicly available dataset that contains APT attack traces injected by a red team over the course of two days. This dataset solely includes event logs from windows machines and encompasses over 17 billion events, with only 0.0016\% of events being malicious. While the dataset's class imbalance may lead to less accurate predictions of various attack scenarios, its extensive log volume makes it an ideal candidate to evaluate the performance of language models in detecting APT, compared to other APT sources.

The contributions of this paper are briefly summarized as follows:
\begin{enumerate}
    \item We introduce a novel approach to process system logs, creating a system provenance graph-based attack detection framework that leverages transformer-based language models for APT detection.
    \item We evaluate the proposed framework using two recent APT datasets, comparing its performance against other transformer-based language models and LSTM. Results demonstrate that our approach achieves the highest accuracy and F1-score in detecting APTs.
    \item We establish that our approach is generalizable and outperforms LSTM by a notable margin as the training data size increases, highlighting its superiority in handling larger and more complex log datasets.
\end{enumerate}

This paper is organized as follows. In sections \ref{sec:bg} and \ref{sec:rw} we discuss the background related to the transformer architecture, advanced persistent threats and previous works in this field respectively. We present our approach and methodology in section \ref{sec:oop}. The dataset construction and further analysis are detailed in sections \ref{sec:ctd} and \ref{sec:eda}. Our experimentation and results are presented in section \ref{sec:exp}. Finally, our conclusion is presented in section \ref{sec:conc}.

\section{Background}\label{sec:bg}

In this section, we briefly overview the Transformer architecture and the characteristics of an APT attack.
\subsection{Transformer}
The Transformer architecture is a sequence-to-sequence network that relies solely on attention mechanisms, eliminating the need for recurrent and convolutional components. In recent studies, the Transformer has demonstrated exceptional performance, surpassing many RNN-based models in Neural Machine Translation. The Transformer consists of two main components: an encoder and a decoder, each comprising stacks of multiple identity blocks.

Within the encoder, each block consists of two subnetworks: a multi-head attention mechanism and a feed-forward network. Within the decoder, each block includes an additional masked multi-head attention mechanism compared to the encoder block. Both encoder and decoder blocks are equipped with residual connection and layer normalization, contributing to the overall stability and effectiveness of the Transformer architecture.

Transformer employs self-attention as a fundamental mechanism to process input sequences. Self-attention allows the model to focus on different elements of the input sequence when generating the output at each position. It is a key factor in the Transformer's ability to capture long-range dependencies and relationships between elements in the input sequence.

\subsection{Language Models}
Language models (LMs) are a subset of machine learning models trained to predict the next word in a sequence, given the words that came before it. At their core, LMs capture the underlying structure and patterns in the language, enabling them to generate coherent and contextually relevant text.

The development of deep learning has paved the way for more sophisticated LMs, with the transformer architecture being particularly transformative. Introduced by Vaswani et al.\cite{vaswani2017attention} in 2017, transformers utilize self-attention mechanisms to weigh the relevance of each word in a sequence when predicting the next word. This allows for greater contextual understanding, making transformers particularly adept at capturing long-range dependencies in text.

One significant implementation of the transformer architecture is BERT (Bidirectional Encoder Representations from Transformers) introduced by Google in 2018. Unlike traditional LMs that predict words in a unidirectional manner (either left-to-right or right-to-left), BERT is trained to predict words in a bidirectional manner, considering both the left and the right context in all layers. This deep bidirectional understanding enabled BERT to achieve state-of-the-art results on a range of NLP tasks.

Following BERT's success, Facebook introduced RoBERTa (A Robustly Optimized BERT Pretraining Approach), which is a variant of BERT. RoBERTa modifies key hyperparameters in BERT, training the model longer and on more data, and removing the next-sentence pretraining objective, which further improved performance on several benchmarks.

While LMs have been predominantly applied in natural language processing tasks such as translation, summarization, and question-answering, their ability to recognize patterns and nuances in large datasets can be harnessed in other domains, including cybersecurity. In the context of detecting Advanced Persistent Threat (APT) attacks, transformer-based LMs like BERT and RoBERTa can be utilized to understand and identify suspicious patterns in textual data, such as logs or network traffic metadata, thus providing a novel approach to cyber threat detection.

\subsection{Advanced Persistent Threat}
Advanced persistent threats, or APTs, are a type of cyber attack that is characterized by its stealthiness and ability to remain undetected for long periods of time. APTs are typically targeted attacks, meaning that they are directed at specific organizations or individuals for the purpose of stealing sensitive data or disrupting operations. These attacks are often carried out by nation-states or other well-funded and skilled cybercrime groups, and they may involve the use of custom malware and other advanced tactics. One of the key features of APTs is their ability to adapt and evolve over a longer period of time, making them difficult to detect. While some threat actors work alone, multiple government authorities such as the Cybersecurity and Infrastructure Security Agency (CISA) have linked attacks to APT groups—with some having ties to specific nation-states who use them to further their country’s interests. According to APT trends report Q3 2022, APT campaigns are very widely spread geographically, expanding into Europe, the US, Korea, Brazil, the Middle East and various parts of Asia.

% APT attacks exhibit distinct characteristics, including their targeted attack and well resourced attackers. APT attacks are tailored to specific targets, rendering the learning from a single or a small number of targets less effective. An effective APT detection approach should focus on objectives rather than individual targets, as the objectives of an APT attack campaign for an organization remain specific, regardless of the target. APT actors are typically skilled hackers working in a coordinated manner, possessing both financial and technical resources. As a result, they can sustain prolonged attacks, take advantage of zero-day vulnerabilities and attack tools, and pursue multiple short-term and long-term goals. They maintain their presence undetected for extended periods, persistently attacking their targets. They adapt their strategies and actions, making repeated attempts until they succeed in achieving their purpose. APT actors may employ zero-day exploits to avoid detection based on known signatures and use encryption to obfuscate network traffic. Consequently, an effective APT detection method should not heavily rely on event content due to potential encryption. Furthermore, it should avoid overly relying on the behavior of normal users, as this could hinder its ability to distinguish benign behavior from both normal and malicious users, thereby limiting its capacity for early detection.

% \textcolor{blue}{\subsection{System Provenance Graph}
% }

\section{Related Work}\label{sec:rw}

In this section, we discuss some of the state-of-the-art techniques and approaches for detecting attacks, with a specific focus on advanced persistent threats (APTs). Our work occupies the intersection of several domains- provenance-graph based APT detection, APT detection using machine learning techniques, applications of transformer-based language models in the field of cybersecurity, and log data analysis for threat detection. Hence, we position LogShield within the context of prior research in these realms. \\

\begin{figure}[htbp]
\centerline{\includegraphics[width=0.5\textwidth]{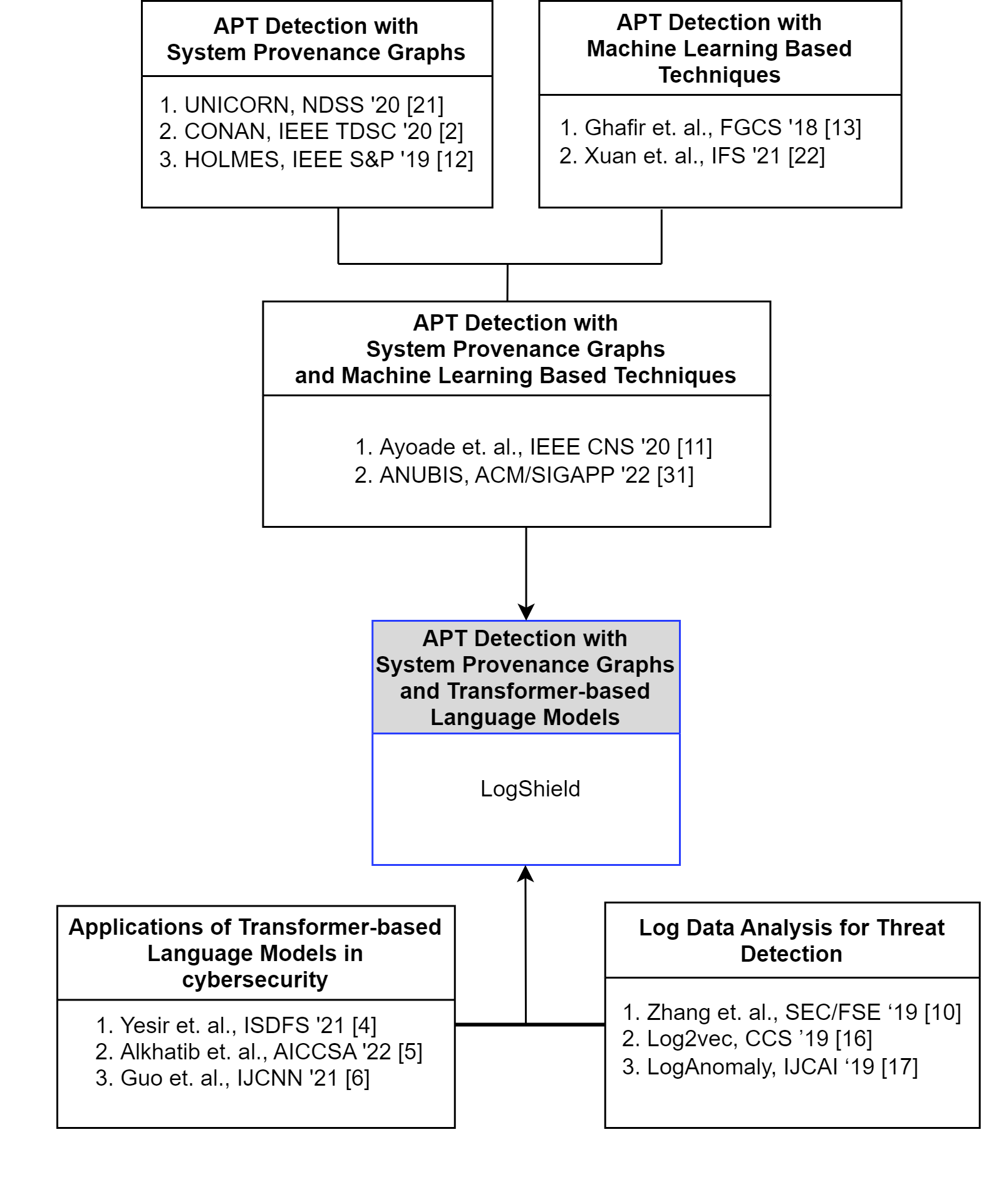}}
\caption{Locating the position of our work in a brief part of the history of APT attack detection}
\label{rel}
\end{figure}

\begin{figure*}[htbp]
\centerline{\includegraphics[width=1\textwidth]{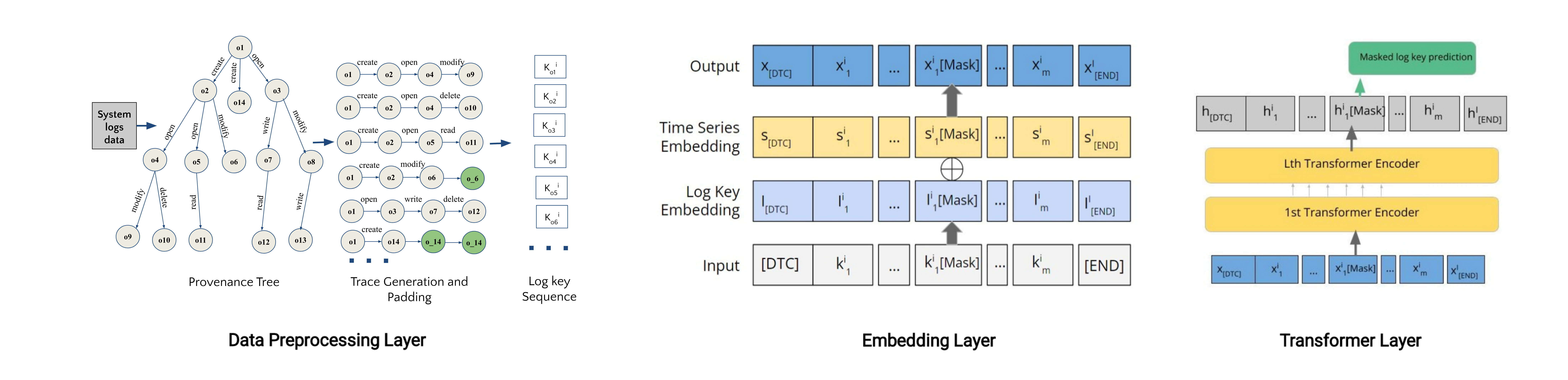}}
\caption{Overview of Our Approach}
\label{fig1}
\end{figure*}

\textbf{Provenance Graph-based APT Detection.} There have been significant prior efforts exploring the use of data provenance for APT detection. Milajerdi et al. \cite{b8} proposed HOLMES, a graph-based method for detecting APT attacks which utilizes a set of manually crafted rules to describe various APT information flows in an attack provenance graph. However, the manual rule generation process has limitations in recognizing zero-day APT attacks. Han et. al. \cite{unicorn} introduced UNICORN, an anomaly based APT detector that effectively leverages data provenance analysis by using a graph sketching technique. However, UNICORN suffers the limitation of having a high false-positive rate in some datasets. Xiong et. al. \cite{xiong2020conan} presented CONAN, a state-based detection framework built upon provenance graphs, which helps to detect APTs with constant and limited memory usage and high efficiency, though the state-based approach can also struggle in identifying zero-day APT attacks.\\

\textbf{APT Detection Using Machine Learning Techniques.} Applying Machine Learning techniques to detect attack patterns is a relatively new idea whereas traditional approaches included either rule-based or signature based methods for attack detection, performance of which were unsatisfactory when detecting zero day APTs. Ghafir et. al. \cite{b9} introduced a machine learning-based correlation analysis method for APT detection. Similarly, Xuan et. al. \cite{Xuan} proposed another ML-based approach using a multi-layer analysis technique. However, their approach demonstrated limitations in effectively detecting novel attacks. In other recent studies, various deep learning based methods have been explored for the detection of APT attacks \cite{b10,b11,b12,b13,b6}. Recurrent neural networks, notably Long Short-Term Memory (LSTM) or Gated Recurrent Unit (GRU), have been extensively utilized in multiple approaches to model normal log sequences \cite{b6,b14}. Machine learning based approaches include supervised and unsupervised deep learning to detect attack sequences. Historically, supervised methods have been employed to tackle the log anomaly detection problem. For example, Liang et al. \cite{b19} took a support vector machine (SVM) method to detect attacks where labels for both benign and malicious samples were available. Various unsupervised learning methods have also been proposed. Xu et al. \cite{b20} suggested using Principal Component Analysis (PCA) assuming distinct sessions in log files are identifiable through session-ids attached to log entries. Additionally, Lou et al. \cite{b21} introduced Invariant Mining (IM) to extract linear relationships among log events from log event count vectors. \\

\textbf{APT Detection Using System Provenance Graphs and Machine Learning-based Techniques.} Our work most closely aligns with prior works in this area. For APT detection, Ayoade et al. \cite{b7} presented a deep learning-based approach where they extracted features from a provenance graph and then used those features to train an online adaptive metric learning model. In another work, ANUBIS \cite{anjum2022anubis}, 
% a provenance graph was constructed from the DARPA OPTC dataset and a Bayesian-Neural Network was leveraged to predict APTs. The main focus of ANUBIS was generating a framework which could explain the predictions of its model to cyber-threat responders. \\
representation learning of event sequences extracted from provenance graph was performed using LSTM network followed by a Bayesian Neural Network classifier for APT detection. \\
% While having room for further improvement, these works demonstrated higher accuracy scores compared to the contemporary state-of-the-art, showcasing the potential for future research advancements in this direction.\\

\textbf{Applications of Transformer-based Language Models in the Field of Cybersecurity.} While no prior works have utilized transformer-based language models for APT detection, in the realm of cybersecurity, there is a growing body of research utilizing transformer-based language models for various tasks. Yesir et al. \cite{yesir2021malware} successfully applied BERT and fastText for malware classification, achieving satisfactory results. More recently, Alkhatib et al. \cite{alkhatib2022can} introduced CAN-BERT, a framework that utilizes BERT for intrusion detection. Similar to our work, Guo et al. \cite{b10} employed the BERT model to capture log sequence information and introduced two self-supervised tasks for model training. However, logBERT relies on a log key parser to convert events in raw log files to log keys.

\textbf{Log Data Analysis for Threat Detection.} Several studies have utilized natural language processing (NLP) techniques to analyze log data, treating logs as natural language sequences. Zhang et al. \cite{b13} employed the LSTM model along with TF-IDF weighting for predicting anomalous log messages. Similarly, LogRobust \cite{b6} and LogAnomaly \cite{b12} integrated pre-trained word vectors to train a Bi-LSTM model for learning log sequences.\\

To the best of our knowledge, our work is the first attempt to leverage the prowess of transformer-based language models for detecting APT attacks. In addition, our approach stands apart from previous methods as we have added customized embeddings on sequence of events generated from provenance graphs, and used transformer architecture to employ self-attention along with log-key cross entropy loss function to enrich the separation of benign and malicious event sequences, leading to significant improvements in APT detection performance.
% We have also conducted our experiements using the latest DARPA OPTC dataset, which has not been used in most of the prior works.

\section{Overview of our approach}\label{sec:oop}
In this section, we introduce LogShield, a framework inspired by Transformer architecture for detecting APT attacks. Logshield has been trained in a supervised manner with customized objective function and embedding layers that learns the separation between benign and malicious event sequences. Our framework utilizes a host-based approach for generating event sequences from provenance graphs and detecting APT attacks. The unique characteristics of APTs, such as ``low-and-slow'' attack pattern and the use of zero-day exploits, are taken into consideration during specific preprocessing. Our approach has two parts: (i) learning log representations from a large log file, and (ii) accurately detecting APT attack patterns. Figure 1 provides a detailed view of this process, which includes: (1) generating log sequences from a provenance graph in the data preprocessing stage, (2) encoding log strings and time in the embedding layer, and (3) learning encoded log representations in the transformer layer.

\subsection{Data Preprocessing Layer}
The layer in question allows for the examination of events that occurred at different times by creating a provenance graph from raw data. Data provenance can be used to model various types of event sequences, such as system logs. While creating a provenance graph, a set of properties (such as object, action etc.) are extracted from each record in the log data to be included in each node of the graph. Through analyzing causality relationships within the provenance graph, a clear understanding of the system's behavior can be gained.

After generation of the provenance graph, we generate event traces. An event trace is a sequence of events that are connected by parent-child relationships. A node in the provenance graph is selected randomly and the path starting from the node to the leaf is selected. Since different paths might have different lengths (which was found to be 31 events per trace on average for the DARPA-OPTC dataset), all the event traces are padded with an offset to make them of exactly same length. As shown by several prior works\cite{b14,b21,b22,b23}, a convenient way to use log data is to generate a more structured representation by storing key properties that preserve causality, while discarding unnecessary information. The resulting representation is referred to as log keys.

\subsection{Embedding Layer}
At this stage, the event traces have been encoded to contain information about each type of event, and on which object the action is being performed upon. Additionally, they possess the time difference information from parent to child event.

\subsubsection{Log Embedding}
A log embedding is a vector representation for each event trace. For each event, we already have its representative object-action pair. We use it to encode an event into natural language text, such as “File Read” or “Process Create” which is then passed on to the encoder architecture of Transformer. Each object-action pair in a sequence will be converted to a token and a sequence will be differentiated from other sequences by a special token. We appended the [DTC] token to mark the start of a sequence and [END] token to denote the end of a sequence, before passing it to the token embedding layer. Some of the tokens are randomly masked by the token embedding layer for the training purpose. BERT and the language models derived from it uses masked language modeling where a masked token is predicted in the output sequence. MLM consists of giving BERT a sequence and optimizing the weights inside BERT to output the same sequence on the other side.

\subsubsection{Temporal Embedding}
We have the time differences for each event from its parent and leveraging this information, we use a sliding window technique to divide the log keys into multiple time segments. In BERT or any other language model, the original input is texts and the temporal information between words is not of much value, unlike our case, where timestamps of each event is very significant. Let's think of two child events: c1 and c2 generated from the same parent. The two events c1 and c2 were not triggered at the same point in time and therefore they cannot be handled similarly. This temporal information becomes handy while detecting APT attacks due to their slow and steady nature. We have introduced temporal embedding layer to differentiate between these type of events.

Due to the nature of APT attacks, the attack sequence is executed slowly and might be kept hidden for a longer amount of time. In our approach, the Temporal Embedding layer leverages the sliding window \textit{w} to divide the log keys into multiple time slots where each log key is placed into one slot. We also divide the value of each time slot by the total number of time slots to ensure the time slot of log keys never surpasses 1. For example, given the sequence of pairs: \textit{(k1, 1s), (k2, 1.5s), (k3, 2.5s), (k4, 5s)}, where the first element of each pair denotes log key, and the second element denotes its corresponding timestamp. If\textit{ }w is defined to be 2 seconds, then the first two pairs, i.e., (k1, 1s) and (k2, 1.5s) are placed into the first slot of time, i.e., 1/3. With the same token, the third and fourth elements in the sequence are placed into the second and the third time-slot, i.e., 2/3 and 3/3, respectively.

\subsection{Transformer Layer}
Transformer, an encoder-decoder architecture that utilizes self-attention mechanisms. In our approach, the transformer layer uses multiple self-attention layers and pointwise fully connected layers that are stacked on the top of each other. It is used to learn the contextual relations among log keys in a sequence. For example, a shell process \textit{sh1} triggers two different shell processes \textit{sh2} and \textit{sh3} at two different timestamps. It also triggers a \textit{File Create} process that creates two files \textit{f2} and \textit{f3} respectively to be used by \textit{sh2} and \textit{sh3}. It is obvious from the context that the temporal embedding of \textit{sh2} and \textit{f2} and that of \textit{sh3} and \textit{f3} should be similar. This similarity score is captured by transformer architecture. The temporal embeddings in the Embedding Layer allows self-attention mechanism to quantify the embedding changes at any given time in the latent space via calculating the cosine similarity.

In our approach, each attention head operates on the output of embedding layer, i.e.,
$X^{i}=\{{x^{i}_{[DTC]},x^{i}_{1},..., x^{i}_{[MASK]j},..., x^{i}_{m} , x^{i}_{[END]}}\}$ where $X^{i}\in X$ is the input to the embedding layer, which is a vector encoded to jointly capture information of different aspects at different positions over the input log sequence. We will refer to $X^{i}$ as log key henceforth. The transformer layer outputs the sequence $H^{i} = \{{h^{i}_{[DTC]},h^{i}_{1},..., h^{i}_{[MASK]j},..., h^{i}_{m} , h^{i}_{[END]}}\}$ where $H^{i}\in \mathcal{H}$ , and $H^{i}$ is a representative of the input log key
$X^{i}$, and $H^{i}$ is calculated using a self-supervised objective function explained in the following.
\begin{figure*}[htbp]
\centering
\subfigure[Percentage of Object-Action Pair in Provenance Graph]{
    \includegraphics[width=0.4\textwidth]{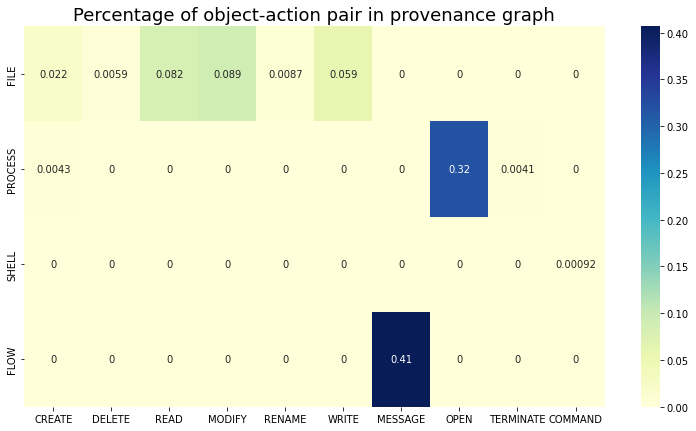}
    \label{fig2}
  }
\subfigure[Comparison of Percentage of Object]{
    \includegraphics[width=0.4\textwidth]{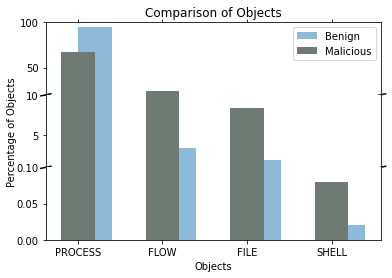}
    \label{fig3}
  }
\caption{Comparison of Percentages of Object-Actions in DARPA OPTC}
% \label{fig2}
\end{figure*}
\subsubsection{Log Key Cross-Entropy Loss}
This objective function is employed to capture context information of encoded log sequences in $K^{i}$. The log keys in each sequence are masked using a specific token [MASK] randomly. Since APT attacks usually exhibit low and slow attack patterns, a system that has been compromised can work similarly to an unattacked system. Thus, the richer contextual information extracted from log key sequence, the better the separation between benign and malicious event sequences will become.

During the training process, the model is expected to learn how to predict masked log keys accurately. The purpose is to encode the prior knowledge of benign log key sequences. To this end, we take the output of the $l$-th encoder in the transformer layer, i.e., $H^{i}$. Then, each masked log key $h^{i}_{[\text{MASK}]j}$ is fed into a softmax function which outputs a probability distribution over the entire set of log keys in the vocabulary $\mathcal{V}$. The highest probability determines the log key in $\mathcal{V}$ that the masked element $h^{i}_{[\text{MASK}]j}$ belongs to. The following equation explains this process where $\hat{k}^{i}_{j}$ indicates the predicted log key for the $j$-th masked element in $H^{i}$.

\begin{center}
    $\hat{k}^{i}_{j} = \text{Softmax}(W_{c}h^{i}_{[\text{MASK}]j} + B_{c})$
\end{center}

Where $W_{c}$ and $B_{c}$ are trainable parameters.

Finally, we adopt the cross entropy loss function to examine how well it could predict the masked log keys, with respect to the corresponding log keys in the same log key sequence.
\begin{center}
    $Loss = -\frac{1}{|\mathcal{H}|}\sum_{i=1}^{|\mathcal{H}|}\sum_{j=1}^{|\hat{K}^{i}|}k^{i}_{j}\log{(\hat{k}^{i}_{j})}$
\end{center}
Where $|\mathcal{H}|$ and $|\hat{K}^{i}|$ indicate the number of input log key sequences and the number of masked log keys in the $i$-th sequence, respectively.

\section{Constructing The Dataset}\label{sec:ctd}
Our experimental datasets includes Darpa OpTC and DARPA Transparent Computing Engagement 3 (TC3). In this section, we present our methodology of constructing the dataset.\\

From the DARPA OPTC Dataset, the log data was collected from 3 Hosts, namely  SysClient0201, SysClient0321 and SysClient0501. Similarly, from the DARPA TC E3 dataset, data of the host TA1-Cadets was collected. The process begins by creating a provenance graph, from the raw data in the log file. This allows for the causal connection of system events even when they occur at different times. The structure of the provenance graph is similar to an n-ary tree. A set of properties are extracted from each record in the log file to be included in the nodes of the tree. The more properties included, the more comprehensive and generalized the model will be. For our provenance tree, we included the following properties - the id of the actor initiating an event, the type of the event (object), the action that was performed (create/modify) and the time difference between the parent event and the child event. 
% This structure of the provenance tree provides valuable information about the context, causality, and neighborhood of a specific system event. 
Each node of the provenance tree had a unique parent node and multiple child nodes. The parent-child relationship is defined by an event triggering another event.

% In our case, the child event is the first event in the dataset, ordered by timestamp, that has the actor id of the parent as its object id. Each node had one of four types of objects on which an action was performed upon - File, Process, Shell and Flow. Each node also had an action associated with it among ten action types, e.g : create, delete, read, modify, rename etc. We associated each type of event with a unique combination of numerical floating point value to represent the object-action pair.

By following Alg \ref{alg:1}, the provenance tree was created from the data, which contained encoded 4 tuples for all nodes starting from parent to child. Since each child has a unique parent but not the other way around, a back provenance tree (a tree containing child to parent mapping) was created by following Alg \ref{alg:2}. That way, we could walk from leaf to root and generate sequence of events known as event traces (Alg \ref{alg:3}). Thus, a collection of benign and malicious event traces was achieved. 

 \RestyleAlgo{ruled}
\begin{algorithm}[]
\caption{Generate Tree $T$}\label{alg:1}
\SetAlgoLined
\SetKwInOut{Input}{Input}\SetKwInOut{Output}{Output}
\Input{Raw Dataset from Host} 
\Output{ $T$ = (list of $C_i$), where $C_i$ is child node\\
$C_i = (I^d,O,A,t^d)$}
\BlankLine
Generate child to parent mapping dictionary \\$\mathit{C}$$\mathit{P}$: $\mathit{v}$ $\rightarrow$ $\mathit{u}$ where $\mathit{u}$ is the id of the first event in the\\dataset (ordered by timestamp) that has that actorID of \\$\mathit{v}$ as it's objectID\;

\For{each $I^d$}
    {Create node \textit{n}, identifying by its $I^d$\;
    Select Object O and Action A for \textit{n}\;
    Find parent from $\mathit{C}$$\mathit{P}$, find time difference $t^d$ from parent\;
    Add \textit{n} as a child to parent node, find parent node from $\mathit{C}$$\mathit{P}$
    }
    
return $T$\\
\end{algorithm}

\begin{algorithm}[]
\caption{Generate Back Tree $T'$}\label{alg:2}
\SetAlgoLined
\SetKwInOut{Input}{Input}\SetKwInOut{Output}{Output}
\Input{$T$} 
\Output{ $T'$ : $C^n \rightarrow P^n$ and \\
\text{$P^n = (I^d, O, A, t^d)$}}
\BlankLine

\For{each $I^d$}{Store parent node $P^n$\;
    }
    From malicious file, find and separate the malicious entries in tree;
    
return $T'$\\
\end{algorithm}

\begin{algorithm}
\caption{Generate Trace}\label{alg:3}
\SetAlgoLined
\SetKwInOut{Input}{Input}\SetKwInOut{Output}{Output}

\Input{$T'$: $C^n \rightarrow P^n$ and \\
$P^n = (I^d, O, A, t^d)$}
\Output{List of event traces $Q$}
\BlankLine
$Q \gets$ initialize as $\emptyset$\;

\While{Dataset $\neq \emptyset$}{
    Randomly choose event $e$\;
    Trace $q \gets$ empty queue\; 

    \While{$e \neq \text{Null}$}{
        For $e_{id}$, get $(O, A, t^d)$\;
        Replace $(O, A, t^d)$ with textual representation\;
        Add to front of $q$\;
        $e \gets T'[e_{id}]$\;
    }

    Add $q$ to $Q$\;
}

\textbf{return} $Q$\;
\end{algorithm}

\section{Exploratory Data Analysis}\label{sec:eda}
\subsection{DARPA OPTC dataset}
The provenance graph of the Darpa OPTC dataset comprised of 4 types of objects and 8 types of actions, leading to 32 possible object-action pairs. However, some of these pairs were more commonly observed in the data, while others were rare. By utilizing a heatmap shown in figure \ref{fig2}, we could obtain a visual representation of the prevalence of these object-action pairs and gain insights about the data.

Once the event traces were generated, we attempted to compare the characteristics of malicious and benign traces. By analyzing figure \ref{fig3}, it was observed that the proportion of FLOW, FILE and SHELL objects was higher in the malicious traces, while the PROCESS object was more prominent in the benign traces. Similarly, the actions MESSAGE, MODIFY, RENAME, DELETE, and COMMAND can be regarded as more suspicious due to their higher frequency in malicious actions.

\begin{figure*}[htbp]
\centering
\subfigure[DARPA OPTC]{
    \includegraphics[width=0.4\textwidth]{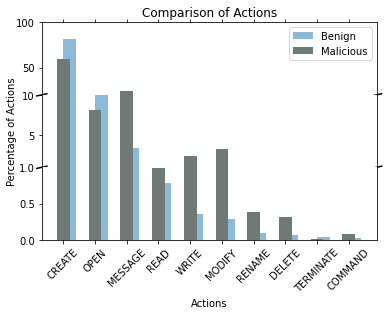}
    \label{fig:action}
  }
\subfigure[DARPA TC E3]{
    \includegraphics[width=0.4\textwidth]{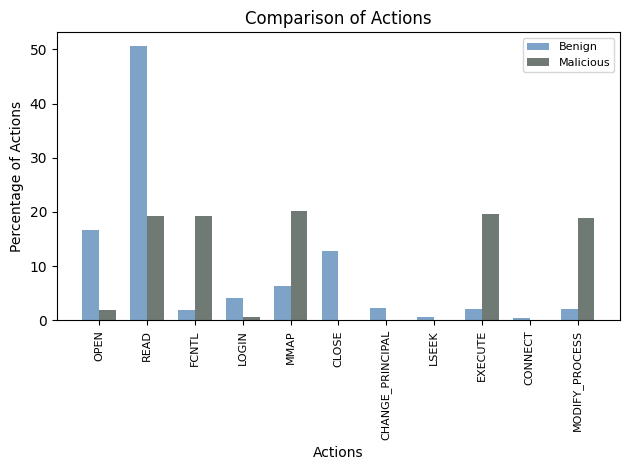}
    \label{fig:action2}
  }
\caption{Comparison of Percentage of Actions}
\label{fig4}
\end{figure*}

By examining the time differences between events in the dataset, another observation that could be made is that malicious actions typically exhibit longer time intervals compared to benign actions. This is logical since the DARPA OPTC dataset contains the data of advanced persistent threats, which are characterized by a slow and steady approach.
\begin{figure}[htbp]
  \centering
  \subfigure[Benign]{
    \includegraphics[width=0.2\textwidth]{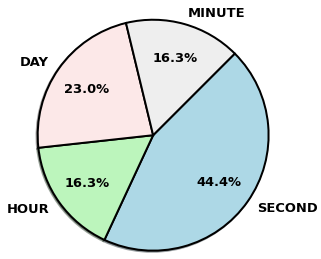}
    \label{fig:first}
  }
  \subfigure[Malicious]{
    \includegraphics[width=0.2\textwidth]{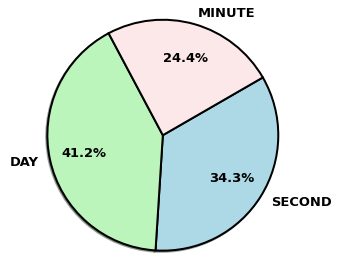}
    \label{fig:second}
  }
  \caption{Comparison of Time Difference Between Events}
  \label{fig:overall}
\end{figure}
Furthermore, analyzing the mutual information (MI) (figure \ref{fig:6}) between the target and each feature reveals that the features ``PROCESS'' and the actions ``CREATE'' and ``OPEN" exhibit the strongest correlation with the target variable.

\begin{equation}
MI(X;Y) = \sum_{y \in Y} \sum_{x \in X} p(x,y) \log \frac{p(x,y)}{p(x) p(y)}
\end{equation}

\subsection{DARPA TC E3 dataset}
In this dataset, there are 17 types of events including READ, WRITE, OPEN, CONNECT, MODIFY, etc. An analysis (figure \ref{fig4}) of the frequency of the events in benign and malicious logs show us that the MODIFY, EXECUTE and MMAP events could be regarded as more suspicious due to their higher frequency in the malicious traces.

\section{Experimentation and Evaluation}\label{sec:exp}
In this section, we aim to evaluate the robustness and effectiveness of our approach. We conducted our experiment on a workstation with an Intel Core i7-10750H running at 2.6 GHz, 16G RAM and 6 cores. We did not use any provenance capturing tool for provenance graph generation except for python libraries. To implement LogShield, a pretrained RoBERTa model was utilized, and the key hyperparameters were adjusted, including the elimination of the next-sentence pretraining objective, as well as training with significantly larger mini-batches and learning rates. Training was conducted using the logs obtained from SysClient0201 and SysClient0501. Considering the computational resource constraints, it was not feasible to train LogShield on the entire Darpa OpTC or Darpa TC E3 dataset.

% \subsection{Experimental Dataset}
% Our experimental datasets includes Darpa OpTC and DARPA Transparent Computing Engagement #3 (TC3). These datasets were specifically introduced by the Defence Advanced Research Projects Agency (DARPA) with the primary objective of comprehending advanced persistent threat (APT) attacks and devising experimental techniques to mitigate such security risks. The datasets encompass an extensive collection of billions of events extracted from an enterprise network, capturing various types of behaviors, both benign and malicious, recorded through network and host-level log telemetry. The datasets' substantial scale and wealth of information render them highly advantageous for training conventional and deep learning models in the realm of APT or anomaly detection.

\subsection{Metrics}
We leverage the widely used metrics, namely Precision, Recall, and F1-score to measure the effectiveness of our approach. The detailed definitions of our metrics are described in the following (TP, FP, FN represent True Positive, False Positive, and False Negative respectively).

\textbf{Precision}: The percentage of correctly detected anomalies amongst all detected anomalies.

\begin{equation}
Pr = \frac{TP}{TP+FP}\label{eq}
\end{equation}

\textbf{Recall}: The percentage of correctly detected anomalies amongst all real anomalies.

\begin{equation}
Rc = \frac{TP}{TP + FN}\label{eq}
\end{equation}

\textbf{F1-Score}: Harmonic mean of precision and recall
\begin{equation}
 F1 = \frac{2 * Pr * Rc}{Pr + Rc}\label{eq}
\end{equation}

% \subsection{Effectiveness of Preprocessing layer}
%  to be written later
\subsection{Effectiveness of the Temporal Embedding layer}
To figure out the effectiveness and robustness of our temporal embedding layer, added on top of our preprocessing layer, we evaluate the performance of existing language models with and without the temporal embedding. By default, language models lack any temporal information pertaining to the textual elements, and instead rely solely on the interplay between word embeddings for text processing. The outcomes of this evaluation, presented in Table-\ref{tab:model-comparison}, demonstrate the notable impact of the temporal embedding layer in enhancing model effectiveness. The models under investigation were subjected to training and testing using distinct collections of host machines. Remarkably, it was observed that when the models were assessed on unfamiliar data from another host machine within the same network, their performance exhibited a slight decline. This observation suggests that our model possesses a degree of generalizability, whereby training on a specific set of host machine data allows for successful application to other hosts, as the individual host being tested does not exert a significant impact on model performance. This generalization holds greater feasibility when employing large language models as opposed to LSTM or other sequence classifiers, as training on the entire dataset necessitates substantially higher computational power.\\
\begin{figure}[htbp]
  \centering
  \subfigure[Objects]{
    \includegraphics[width=0.2\textwidth]{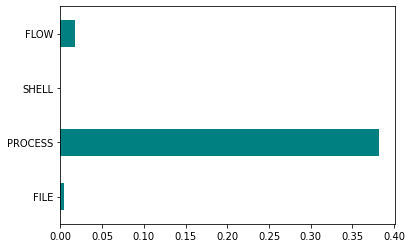}
    \label{fig:first}
  }
  \subfigure[Actions]{
    \includegraphics[width=0.2\textwidth]{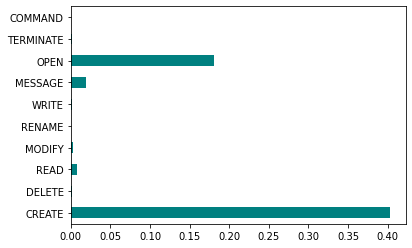}
    \label{fig:second}
  }
  \caption{Mutual Information between features and target}
  \label{fig:6}
\end{figure}
Our experimentation shows that LSTM doesn't improve significantly with temporal embedding like BERT, RoBERTa or ELECTRA. Among the three models we used, RoBERTa has been found to perform better in most of the cases. In the BERT framework, the masking procedure is executed once during data preparation, wherein each sentence undergoes masking in 10 distinct manners. Consequently, during the training phase, the model exclusively encounters these 10 variations of each sentence. On the other hand, the probable reason why RoBERTa might have performed well in event sequence classification is the masking process takes place during training, implying that every time a sequence is included in a minibatch, it undergoes the masking procedure. Consequently, the number of potentially different masked versions for each sentence is not limited as it is in BERT.\\

Finally, Figure-\ref{fig:comp-models} shows how the F1-score improves after integrating temporal embedding for different combinations of trainset and testset.

\subsection{Effectiveness in terms of APT attack detection}
We evaluate the effectiveness of LogShield in terms of APT attack pattern detection. LogShield's performance was evaluated on data collected from SysClient0201, SysClient0321, and SysClient0501. The performance metrics of LogShield are presented in Table-\ref{tab:compare}.

\begin{table*}[t]
\centering
\caption{Comparing LogShield with language models in attack detection}
\begin{tabular}{|c|*{8}{c|}}
\hline
\multirow{2}{*}{\textbf{Model}} & \multicolumn{4}{c|}{\textbf{DARPA OpTC}} & \multicolumn{4}{c|}{\textbf{DARPA TC E3}} \\

\cline{2-9}
& \textbf{Precision} & \textbf{Recall} & \textbf{Accuracy} & \textbf{F1 Score} & \textbf{Precision} & \textbf{Recall} & \textbf{Accuracy} & \textbf{F1 Score} \\
\hline
BERT & 0.89 & 0.91 & 0.95 & 0.90 & 0.85 & 0.89 & 0.88 & 0.87 \\
\hline
% RoBERTa & 0.96 & 0.93 & 0.95 & 0.94 & 0.96 & 0.91 & 0.92 & 0.93 \\
% \hline
LogShield & 0.98 & 0.99 & 0.99 & 0.984 & 0.95 & 0.95 & 0.96 & 0.95 \\
\hline
GPT-2 & 0.92 & 0.92 & 0.94 & 0.92 & 0.90 & 0.94 &  0.91 & 0.92\\
\hline
LSTM & 0.97 & 0.97 & 0.98 & 0.96 & 0.95 & 0.94 & 0.93 & 0.944 \\

\hline
\end{tabular}
\label{tab:compare}
\end{table*}

We also conducted a comprehensive analysis to investigate the ability of LogShield to accurately classify traces that were misclassified by the LSTM model. Through various permutations of trainset and testset combinations, we examined the performance of LogShield in correctly categorizing these misclassified traces. The detailed findings of our analysis are presented in Table-\ref{tab:misclassification}.
\begin{table}[H]
\begin{tabular}{|l|l|c|}
\hline
\textbf{Trained On}            & \textbf{Tested On} & \textbf{\begin{tabular}[c]{@{}c@{}}Correctly Classified by logShield \\ (Misclassified traces by LSTM)\end{tabular}} \\ \hline
\multirow{3}{*}{SysClient0201} & Sysclient0201      & 503 (2503)                                                                                                           \\ \cline{2-3} 
                               & Sysclient 0321     & 7956 (22529)                                                                                                         \\ \cline{2-3} 
                               & Sysclient0501      & 5455 (22503)                                                                                                         \\ \hline
\multirow{3}{*}{SysClient0501} & Sysclient0201      & 12457 (25503)                                                                                                        \\ \cline{2-3} 
                               & Sysclient 0321     & 6001 (20503)                                                                                                         \\ \cline{2-3} 
                               & Sysclient0501      & 506 (2503)                                                                                                           \\ \hline
Darpa TC E3                    & TA1-cadets         & 427 (2574)                                                                                                           \\ \hline
\end{tabular}
\caption{Count of traces correctly classified by LogShield which were misclassified by LSTM}
\label{tab:misclassification}
\end{table}

\subsection{Relation of Training Data Size with Performance}

\begin{figure}[H]
  \centering
\includegraphics[width=0.5\textwidth]{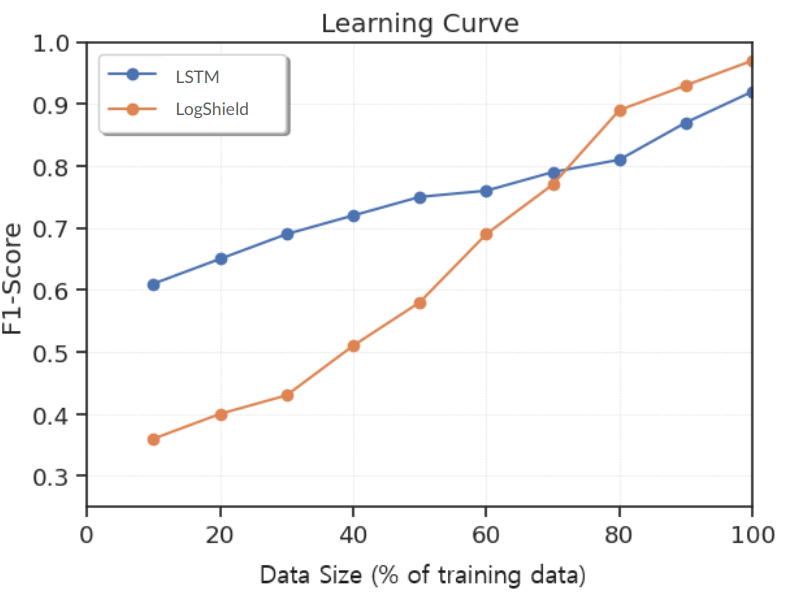}
  \caption{Comparison of Performance of LSTM and LogShield with respect to varying sizes of training data}
  \label{fig:comp-data_size_vs_models}
\end{figure}

\begin{figure*}[htbp]
  \centering
\includegraphics[width=1\textwidth]{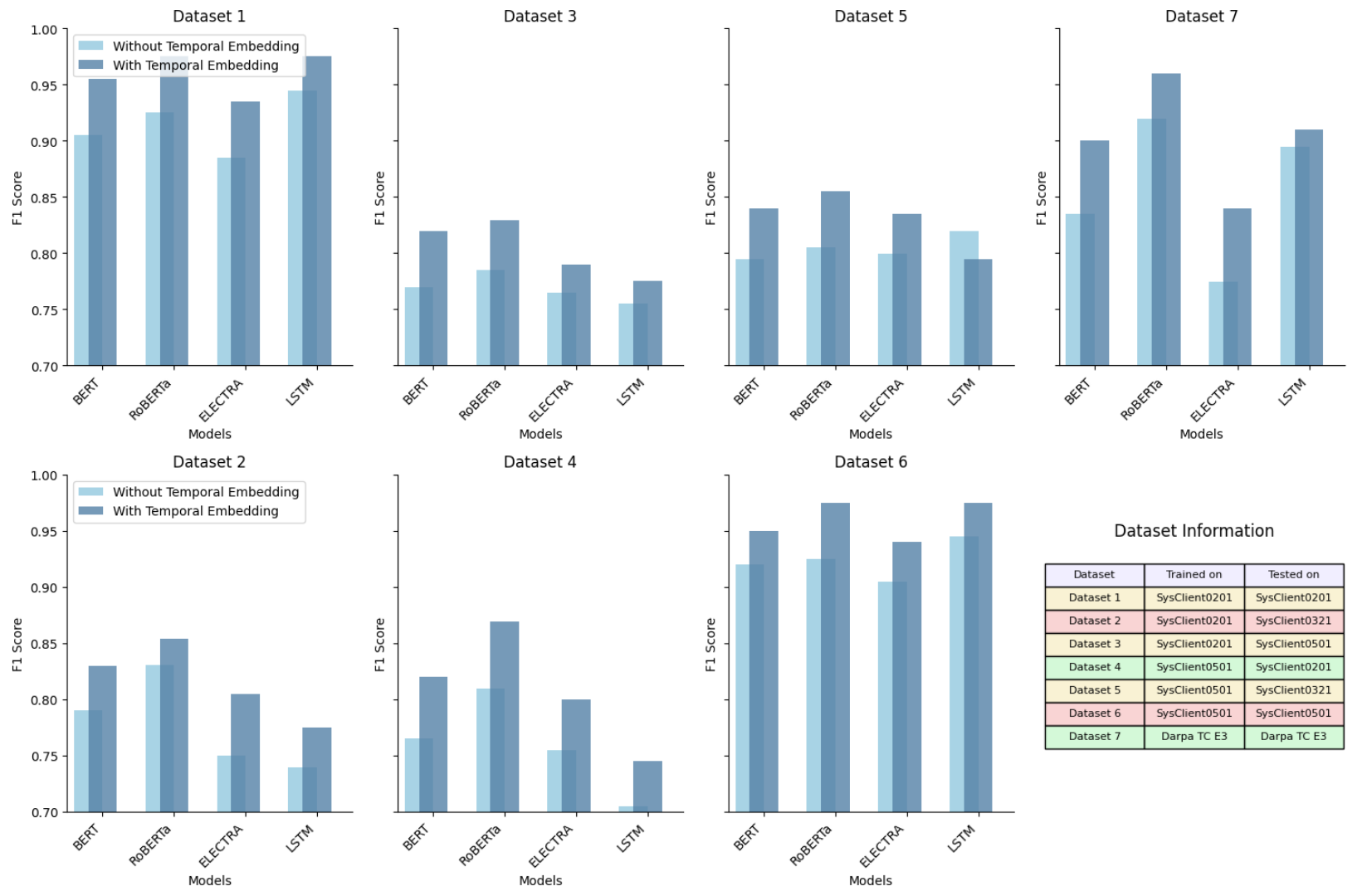}
  \caption{Comparison of Performance of Models with and without Temporal Embedding}
  \label{fig:comp-models}
\end{figure*}

We also evaluated the performance of LogShield and LSTM models with varying sizes of training data. The results, presented in Fig-\ref{fig:comp-data_size_vs_models}, demonstrate that LSTM exhibits superior performance when the training data size is small. However, as the training data size increases, LogShield gradually surpasses LSTM in terms of effectiveness.
Considering the prevalence of substantial system logs in real-life security scenarios, it is reasonable to assume that most attack detection tasks will have access to a considerable amount of training data. Consequently, LogShield emerges as the preferred choice in the majority of cases.

% \section{Discussion}

\subsection{Misclassifications by LogShield}
There are some traces, although few, which were misclassified by LogShield yet correctly classified by LSTMs, as seen in Table-\ref{tab:misclassification-ls}.
\begin{table}[H]
\begin{tabular}{|l|l|c|}
\hline
\textbf{Trained On}            & \textbf{Tested On} & \textbf{\begin{tabular}[c]{@{}c@{}}Correctly Classified by LSTM \\ (Misclassified traces by LogShield)\end{tabular}} \\ \hline
\multirow{3}{*}{SysClient0201} & Sysclient0201      & 45 (2503)                                                                                                           \\ \cline{2-3} 
                               & Sysclient 0321     & 197 (14573)                                                                                                         \\ \cline{2-3} 
                               & Sysclient0501      & 243 (17048)                                                                                                         \\ \hline
\multirow{3}{*}{SysClient0501} & Sysclient0201      & 587 (13046)                                                                                                        \\ \cline{2-3} 
                               & Sysclient 0321     & 369 (14503)                                                                                                         \\ \cline{2-3} 
                               & Sysclient0501      & 65 (2503)                                                                                                           \\ \hline
Darpa TC E3                    & TA1-cadets         & 14 (1147)                                                                                                           \\ \hline
\end{tabular}
\caption{Count of traces correctly classified by LSTM which were misclassified by LogShield}
\label{tab:misclassification-ls}
\end{table}

This can happen for traces that are much shorter than average, as LSTMs are known to be capable of learning effectively from smaller amounts of data. Also, language models may focus more on capturing global patterns and may not fully exploit the local sequential information in some cases, where LSTMs might be more successful.

\subsection{Handling Class Imbalance}
\begin{figure}[H]
\subfigure[DARPA OPTC]{
    \includegraphics[width=0.2\textwidth]{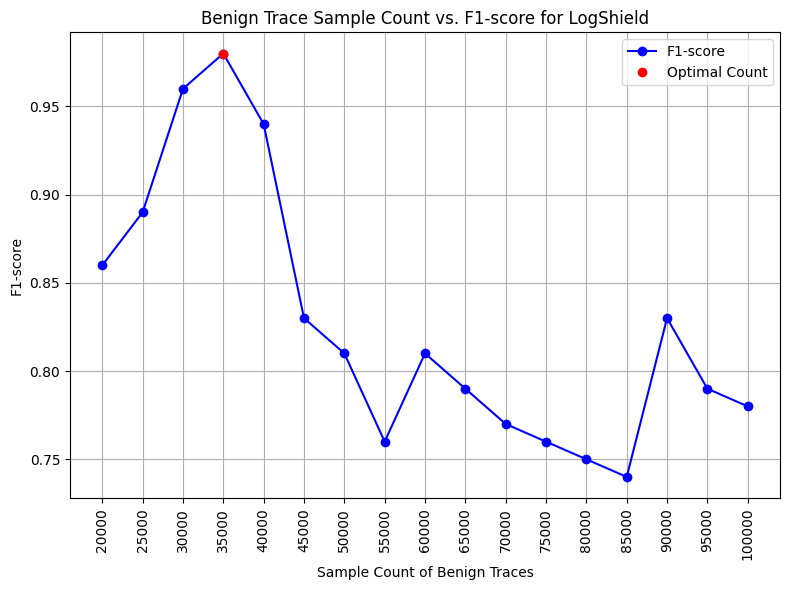}
    \label{fig:ablation1}
  }
\subfigure[DARPA TC E3]{
    \includegraphics[width=0.2\textwidth]{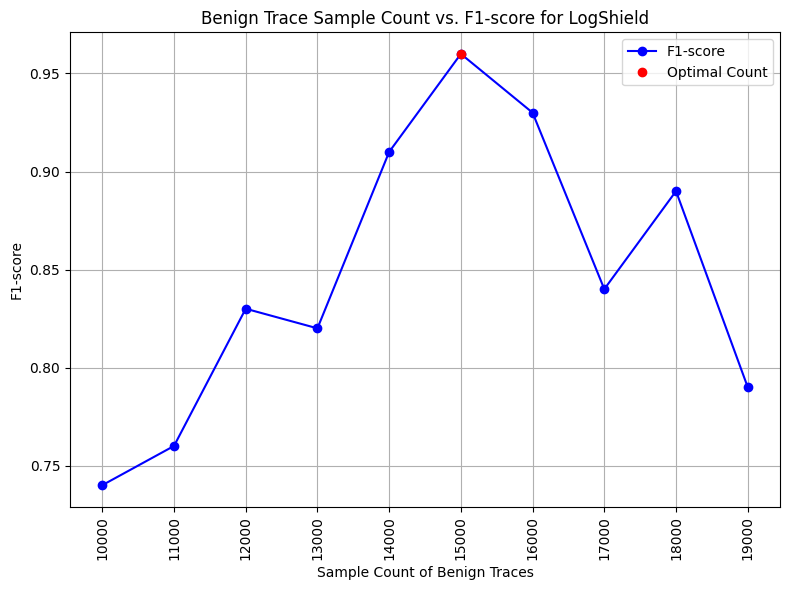}
    \label{fig:ablation2}
  }
\caption{Finding the Optimal Benign Trace Sample Count}
\label{fig:ablation}
\end{figure}

In our analysis of the DARPA OPTC dataset, we extracted 103,329 benign traces and 20,924 malicious traces. For the DARPA TC E3 dataset, we obtained 19,152 benign traces and 9,455 malicious traces. Due to the significant class imbalance, we conducted experiments involving upsampling the malicious traces and downsampling the benign traces. While upsampling the malicious traces did not enhance the performance, downsampling the benign traces resulted in improved model F1-score. Figure-\ref{fig:ablation} illustrates our experimentation process to determine the optimal downsampled count for the benign traces. After finding the optimal count of benign traces for LogShield, the sample count of the traces were kept the same across experimentation with other models.

\subsection{Limitations}
\begin{figure}[htbp]
  \centering
  \includegraphics[width=0.5\textwidth]{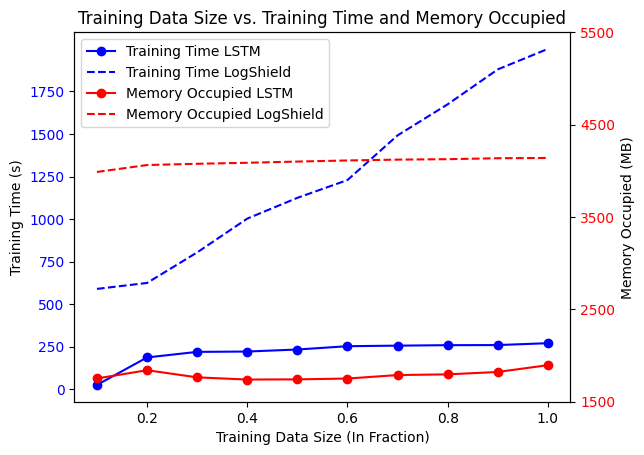}
  \caption{Comparison of Training Time and Memory Occupied for LogShield and LSTM Models with Varying Training Data Sizes}
  \label{fig:time_memory}
\end{figure}

Language models, with their superior performance, outperform LSTMs in terms of accuracy and F1-score, thereby detecting attacks successfully in cases where LSTMs fail. However, it is important to note that this improved performance comes at the expense of increased memory consumption and longer computational time. In Figure-\ref{fig:time_memory}, we illustrate the comparison of training time and memory usage between LogShield and LSTM models using our training data. Considering systems with memory constraints, LSTM models might still be relevant and viable for attack detection applications.

In conclusion, our evaluation demonstrates the superiority of our temporal embedding approach in attack detection, as evidenced by higher precision, recall, and F1-scores across different hosts in the datasets. We also demonstrate that as the training data size increases, LogShield gradually surpasses LSTM in terms of effectiveness. Furthermore, LogShield is able to correctly classify many traces that LSTM fails to detect. We also discussed how we handled the class imbalance of the dataset, as well as the limitation of higher memory consumption of our approach compared to LSTM.

% \section{Future Work}\label{sec:fw}

Our work could be extended in the following directions in future:
\subsubsection{Pretraining LogShield on Larger Dataset}
The DARPA OPTC dataset contains event logs from 1000 hosts, of which we have used the data of only three hosts. There are also a number of similar cyber attack datasets. One similar dataset to OpTC is the Los Alamos National Laboratory (LANL) Unified Host and Network dataset, which records network and host activities from the laboratory over a 90-day period. These additional data could be used to pretrain LogShield to improve its performance.

\subsubsection{Upgrading the Architecture}
The architecture of LogShield might be improved by further modifying the pre-processing layer. More classes of information from system logs can be incorporated to enrich the information in the traces. For example, in the DARPA TC E3 dataset, there were object information of each event such as File Object, Memory Object which was not incorporated in our pre-processing. Additionally, the incorporation of timestamps into the temporal embedding resulted in some loss of information. This aspect could be enhanced by including more precise timing values to improve accuracy and fidelity.
\subsubsection{Improving Prediction}
To improve prediction performance, conducting additional experiments with thorough hyperparameter tuning can be beneficial. Techniques like Grid Search can aid in identifying more appropriate hyperparameters for the model. 

\section{Conclusion}\label{sec:conc}
There have been several successful past efforts in attack detection using deep learning approaches. In this paper, we have explored the effectiveness of transformer based language models in attack detection, particularly advanced persistent threat detection. We have designed a novel approach for processing the system logs from APT datasets, encoding the generated event sequences with multiple embedding layers leveraging transformer architectures. Through extensive experimentation, we have shown that LogShield achieves 98\% F1 score in APT detection. The results substantiate the potential of transformer models in the field of cybersecurity to outperform traditional methods. Looking ahead, our future work will focus on enhancing the architecture by incorporating more comprehensive information from the logs into the embeddings. By harnessing additional contextual data, we aim to further boost the accuracy and robustness of our model, contributing to more effective APT detection strategies.

\section{Acknowledgement}\label{sec:ack}
We express our sincere gratitude to Md. Monowar Anjum and Majid Babaei, whose expertise and support significantly contributed to the completion of this study.

\vspace{12pt}
% \color{red}
% IEEE conference templates contain guidance text for composing and formatting conference papers. Please ensure that all template text is removed from your conference paper prior to submission to the conference. Failure to remove the template text from your paper may result in your paper not being published.

\end{document}